\documentclass[conference,10pt]{IEEEtran}
\IEEEoverridecommandlockouts
\usepackage{amsfonts}
\usepackage{amsmath}
\usepackage{cite}

\usepackage{amssymb}
\usepackage{algorithmic}
\usepackage{graphicx}
\usepackage{textcomp}
\usepackage{xcolor}
\usepackage{amsthm}
\usepackage{verbatim}
\usepackage{balance}
\usepackage{multirow}
\usepackage{booktabs}
\usepackage{url}
\usepackage[switch]{lineno}
\usepackage{soul}
\usepackage[colorlinks, linkcolor=blue, anchorcolor=blue,citecolor=blue]{hyperref}
\soulregister{\cite}7 
\soulregister{\citep}7 
\soulregister{\citet}7 
\soulregister{\ref}7 
\soulregister{\pageref}7 
\soulregister{\texttt}7 
\def\BibTeX{{\rm B\kern-.05em{\sc i\kern-.025em b}\kern-.08em
    T\kern-.1667em\lower.7ex\hbox{E}\kern-.125emX}}
\usepackage[utf8]{inputenc}
\usepackage[ruled,linesnumbered,vlined]{algorithm2e}
\usepackage{threeparttable}
\SetKwInput{KwInput}{Input}
\SetKwInput{KwOutput}{Output}

\newcommand{\todomagenta}[1]{\todoc{magenta}  {[#1]}}
\newcommand{\todoc}[2]{{\textcolor{#1} {#2}}}

\newcommand{\civi}[1]{\todomagenta{Ming: #1}}

\newcommand{\hytt}[1]{\texttt{\hyphenchar\font=\defaulthyphenchar #1}}
\renewcommand{\thefootnote}{}
\usepackage{diagbox}

\begin{document}

\title{Characterizing and Detecting Configuration Compatibility Issues in Android Apps\\
{}
\thanks{\textsuperscript{$*$}Shing-Chi Cheung is the corresponding author of this paper.}
}
\author{\IEEEauthorblockN{Huaxun Huang\textsuperscript{$\dagger$}, Ming Wen\textsuperscript{$\S$}, Lili Wei\textsuperscript{$\dagger$}, Yepang Liu\textsuperscript{$\P$}, Shing-Chi Cheung\textsuperscript{$\dagger$*}}
	\IEEEauthorblockA{\textsuperscript{$\dagger$}\textit{Dept. of Computer Science and Engineering, The Hong Kong University of Science and Technology, Hong Kong, China}}
	\IEEEauthorblockA{\textsuperscript{$\S$}\textit{School of Cyber Science and Engineering, Huazhong University of Science and Technology, Wuhan, China}}
	\IEEEauthorblockA{\textsuperscript{$\P$}\textit{Dept. of Computer Science and Engineering, Southern University of Science and Technology, Shenzhen, China}}
	Emails: \{hhuangas@cse.ust.hk, mwenaa@hust.edu.cn, liliwei@cse.ust.hk, liuyp1@sustech.edu.cn, scc@cse.ust.hk\}
	}

\maketitle

\begin{abstract}

XML configuration files are widely used in Android to define an app's user interface and essential runtime information such as system permissions.
As Android evolves, it might introduce functional changes in the configuration environment, thus causing compatibility issues that manifest as inconsistent app behaviors at different API levels.
Such issues can often induce software crashes and inconsistent look-and-feel when running at specific Android versions.
Existing works incur plenty of false positive and false negative issue-detection rules by conducting trivial data-flow analysis while failing to model the XML tree hierarchies of the Android configuration files.
Besides, little is known about how the changes in an Android framework can induce such compatibility issues.
To bridge such gaps, we conducted a systematic study by analyzing 196 real-world issues collected from 43 popular apps.
We identified common patterns of Android framework code changes that induce such configuration compatibility issues.
Based on the findings, we propose \textsc{ConfDroid} that can automatically extract rules for detecting configuration compatibility issues.
The intuition is to perform symbolic execution based on a model learned from the common code change patterns.
Experiment results show that \textsc{ConfDroid} can successfully extract 282 valid issue-detection rules with a precision of 91.9\%.
Among them, 65 extracted rules can manifest issues that cannot be detected by the rules of state-of-the-art baselines.
More importantly, 11 out of them have led to the detection of 107 reproducible configuration compatibility issues that the baselines cannot detect in 30 out of 316 real-world Android apps.
\end{abstract}


\section{Introduction}
\label{sec:1}

The Android framework provides a flexible XML configuration environment, which is widely used by developers to control Android components' behaviors or even the entire apps, such as defining User Interface (UI) structures of~the apps' layout and declaring the required system permissions,~and so on.

Android continuously evolves to meet different market demands, resulting in
successive releases of thirty
different API
levels since its launch~\cite{androiddevelopers}.
Each API level introduces functional changes to the Android configurations to cater for revised requirements.
Such functional changes can cause the same configuration element in an Android
app to manifest inconsistent behaviors at different API levels. We refer to
such inconsistencies as \textbf{configuration compatibility issues}, which can
lead to poor user experience.
\sethlcolor{yellow}
\begin{figure}[t]
	\centering
	\includegraphics[width=0.5\textwidth]{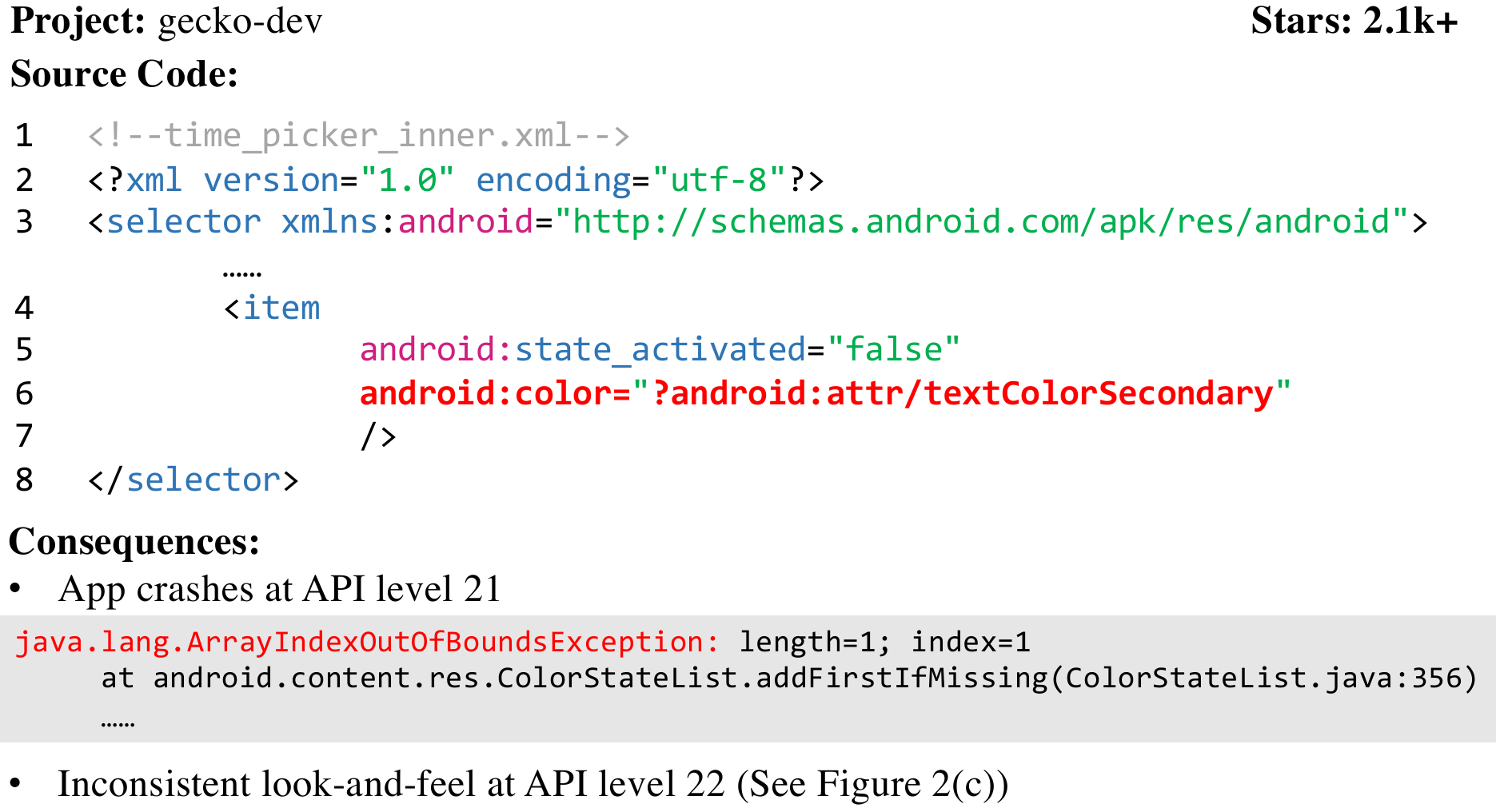}
	\caption{
		%\huaxun{Condition \#3: Revise crash stack trace at API level 21, add inconsistent look-and-feel at API level 22.}
		{A real-world example of configuration compatibility issues reported in \texttt{BUG 1486200} of gecko-dev.}}
	\label{fig:configurationfile}
\end{figure}

Figure~\ref{fig:configurationfile} shows a real-world example of a configuration compatibility issue (i.e. \texttt{BUG 1486200}~\cite{geckoissue}) reported by an open-source Android browser engine project called gecko-dev~\cite{gecko}, which has received 2.1K+ stars in GitHub~\cite{github}.
Lines 3-8 specify a \texttt{<selector>} configuration element to create a color
state list object.
It has a child tag named~\texttt{<item>} containing a set of attributes
such as \texttt{android:color} in Line~6.
{The configuration element triggers a compatibility issue, where the app normally works at an API level $\geq$ 23 but throws \texttt{ArrayIndexOutOfBoundsException} at API level 21 and manifests inconsistent look-and-feel at API level 22. The issue was caused by the different implementations among API levels 21, 22, and 23 to process the attribute value of \texttt{android:color} from XML configuration files in the Android framework (see Section~\ref{sec:motivating_example} for more details).}
In our empirical study, we observe that among the 196 real-world configuration
compatibility issues in 43 popular Android apps, 89 (45.4\%) of them can induce app crashes at certain API levels, while a further 88 (44.9\%) can induce inconsistent look-and-feel across different
API levels. This indicates that, in practice, configuration compatibility issues can induce severe problems in Android apps.

Nevertheless, uncovering configuration compatibility issues caused by such changes is non-trivial.
Specifically, our investigation of 200 top-ranked Android apps found that each of them on average specifies 25,991.6 configuration elements spreading across 663.1 XML files.
Typically, Android apps are designed to run on a range of API levels. It is expensive to design tests to check whether all these configuration elements and their attributes are handled as intended across these diverse API levels.
As a result, configuration compatibility issues can easily be missed by app developers.
Although the related official Android documentation, such as Android Developers~\cite{androiddevelopers} and Android API Differences Reports~\cite{differencereport}, records the information of configuration changes that can result in compatibility issues, such documentation can miss many configuration changes that manifest runtime inconsistencies. 
Even if the relevant changes are documented, the documentation can be overlooked by developers.
Thus, an automatic tool to help detect configuration compatibility issues is helpful.

Existing techniques on detecting software misconfigurations~\cite{rabkin2011static, xu2013not, behrang2015users,xu2016early, dong2016orplocator, chen2020understanding, toman2016staccato, reisner2010using} and Android incompatibilities~\cite{fazzini2017automated,ki2019mimic, wei2016taming, wei2018understanding, wei2019pivot, huang2018understanding, li2018cid, he2018understanding, li2018elegant} are not effective in identifying such configuration compatibility issues and pinpointing the root cause.
For example, the issue between API levels 22 and 23 in Figure~\ref{fig:configurationfile} is triggered by using different API calls with different guarded conditions to load the value of \texttt{android:color} in the XML tag~\texttt{<item>}. %\civi{Cannot follow ``color state list''}
Accurate identification of the root cause requires path-sensitive analysis, which can be expensive.
However, if we consider the existence of a compatibility issue whenever there are implementation differences on how the Android framework processes XML attributes, many issues identified are spurious as many code changes are irrelevant to compatibility issues.
It is non-trivial to analyze code changes that can trigger configuration compatibility issues in the Android framework with the large codebase (i.e., 4M+ LOC at API level 30) and a huge number of code changes in history (i.e., 250 git development branches with changes from 566K+ commits until April 2021).
So far, no prior works have studied the common patterns of code changes in the Android framework that induce configuration compatibility.
To fill such gaps, we conducted an empirical study linking the root causes of real configuration compatibility issues in 43 popular open-source Android apps to the code changes in the Android framework. We found several common code change patterns that can induce configuration compatibility issues in the Android framework (Section~\ref{sec:3}). %\civi{Please use reference.}

Based on the findings, we further propose~\textsc{ConfDroid}, which is a static analyzer encoding the common code change patterns to automatically generate issue-detection rules.
Specifically, \textsc{ConfDroid} performs an intra-class level symbolic execution based on the insight that \textit{common configuration compatibility issues are induced by the code changes within a single class that can process XML attributes in the Android framework}.
In this way, we can greatly reduce the cost of conducting path-sensitive analysis while ensuring comparable accuracy when identifying detection rules.

We implemented \textsc{ConfDroid} based on Soot~\cite{lam2011soot} and ran it on the Android framework code among API levels 21-30 to extract rules for detecting configuration compatibility issues.
The experimental results show that \textsc{ConfDroid} achieved a precision of 91.9\% by successfully generating 282 valid detection rules, which can be reproducible when manifesting runtime inconsistencies in the Android emulators.
Besides, \textsc{ConfDroid} can generate 65 validated rules that are missed by three state-of-the-art baseline methods.
Furthermore, we evaluated the usefulness of \textsc{ConfDroid} for issue detection in the 316 real-world Android apps from F-Droid~\cite{fdroid} and AppBrain~\cite{appbrain}.
Among 65 valid rules that are uniquely returned by \textsc{ConfDroid}, 11 of them have been enabled to generate 107 warnings that can be reproduced to manifest configuration compatibility issues in 30 apps.
So far, 52 warnings have been confirmed, and 51 warnings have been fixed by the developers.
We released the empirical and experiment datasets as well as the \textsc{ConfDroid} artifact in our project website~\cite{confdroid}.

To summarize, our work makes three major contributions.
\begin{itemize}
	\item An empirical research on open-source apps to help understand the common root causes and patterns of configuration compatibility issues.
	\item A technique named \textsc{ConfDroid} that can generate rules to facilitate an automatic detection of configuration compatibility issues in Android apps.
	\item An empirical evaluation showing that (1) \textsc{ConfDroid} outperforms existing approaches on generating new detection rules with high precision, and (2) rules extracted by \textsc{ConfDroid} can be successfully applied for detecting previously-unknown configuration compatibility issues.
\end{itemize}

{\Huge }
\section{Background \& Motivation}
\label{sec:2}
\sethlcolor{yellow}
\begin{figure}[t]
	\centering
	\includegraphics[width=0.5\textwidth]{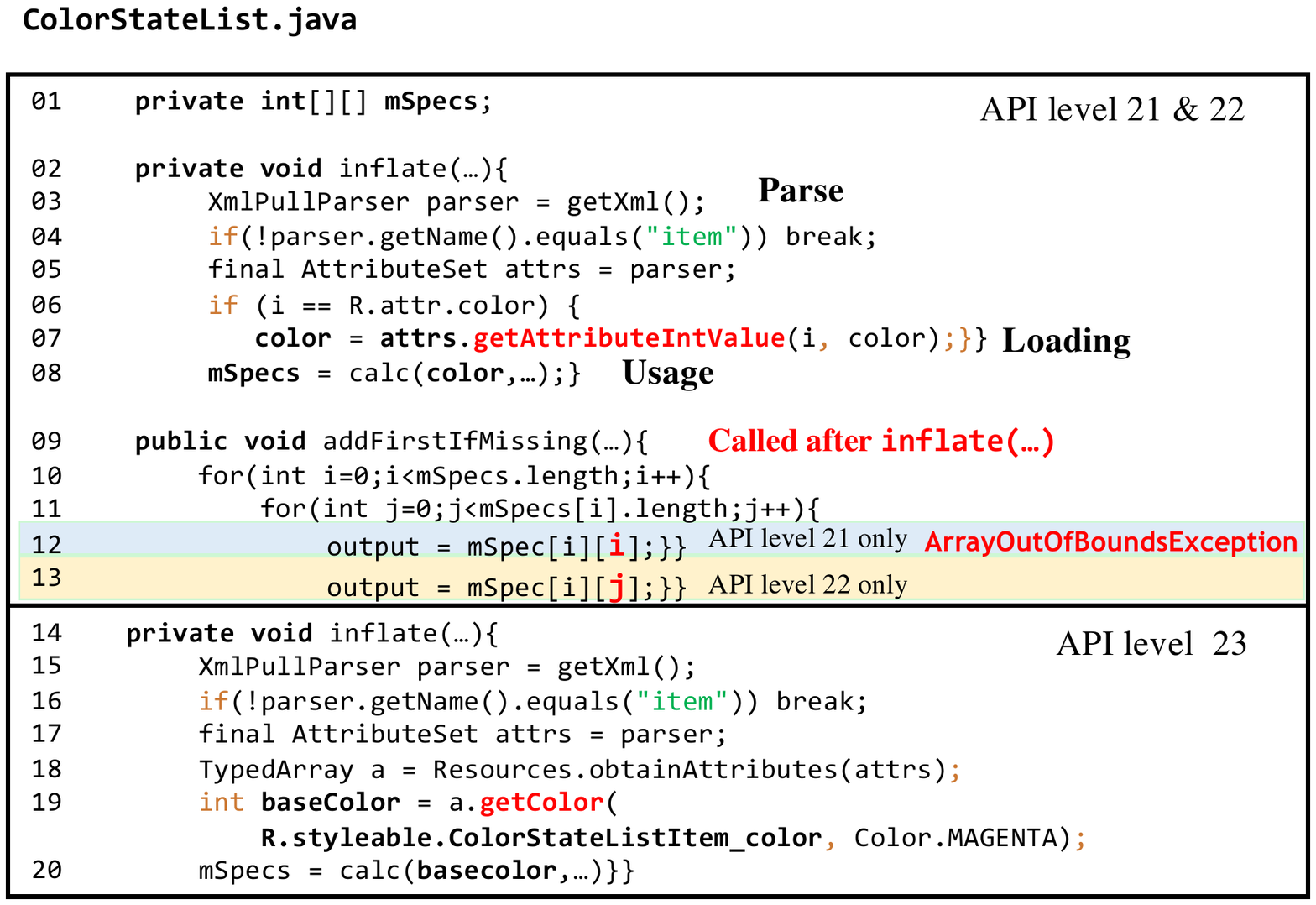}
	\caption{
		{Code snippets adapted from \texttt{ColorStateList} in the Android framework to process the attribute \texttt{android:color} among API level 21, 22 and 23.}}
	\label{fig:frameworkprocess}
	
\end{figure}

\subsection{Processing XML Configuration Files in Android}

XML configuration files are an indispensable part of Android projects.
In general, there are two types of XML configuration files: (1) manifest files (\texttt{AndroidManifest.xml}) that provide the essential runtime information for Android apps, (2) resource XML files (located in the \texttt{/res/} folder) that are commonly used to define an app's user interfaces~\cite{xu2018deeprefiner}.
The Android framework provides a flexible configuration environment that can accept substantially different attributes (e.g., 1,435 different attributes at API level 30).
As the Android framework evolves, compatibility issues arise from inconsistent handling of these attributes and their values among different API levels. 

A popular Android app typically consists of a few hundred XML configuration files. Each is processed as the following steps by the Android framework.

\textbf{Parsing.}
The Android framework uses \texttt{XmlPullParser} to parse all the XML tags in configuration files and return an \texttt{AttributeSet} or \texttt{TypedArray} object for each XML tag.
In particular, \texttt{AttributeSet} and \texttt{TypedArray} are two classes defined in the Android framework to store each XML attribute value as a key-value pair.
The framework provides APIs to trigger the parsing process and returns a newly created \texttt{AttributeSet} or \texttt{TypedArray} object, such as \hytt{getXml()} in Line 1 of Figure~\ref{fig:frameworkprocess}.

\textbf{Loading.}
\texttt{AttributeSet} and \texttt{TypedArray} define APIs (e.g., \texttt{AttributeSet\#getAttributeIntValue()} and \texttt{TypedArray\#getBoolean()}) that take configuration attributes as parameters and return the corresponding attribute values in predefined formats. We refer to these APIs as \textbf{configuration APIs}. In the loading step, the attribute values in the \texttt{AttributeSet} and \texttt{TypedArray} objects are loaded to some program variables using the configuration APIs. 

\textbf{Usage.} The variables are processed by the Android framework to represent the app's runtime behavior.

Configuration compatibility issues can occur when there are inconsistent implementations in the above steps
 among the different Android API levels.

\subsection{Motivating Example}
\label{sec:motivating_example}
\sethlcolor{yellow}
\begin{figure}[t]
	\centering
	\includegraphics[width=0.5\textwidth]{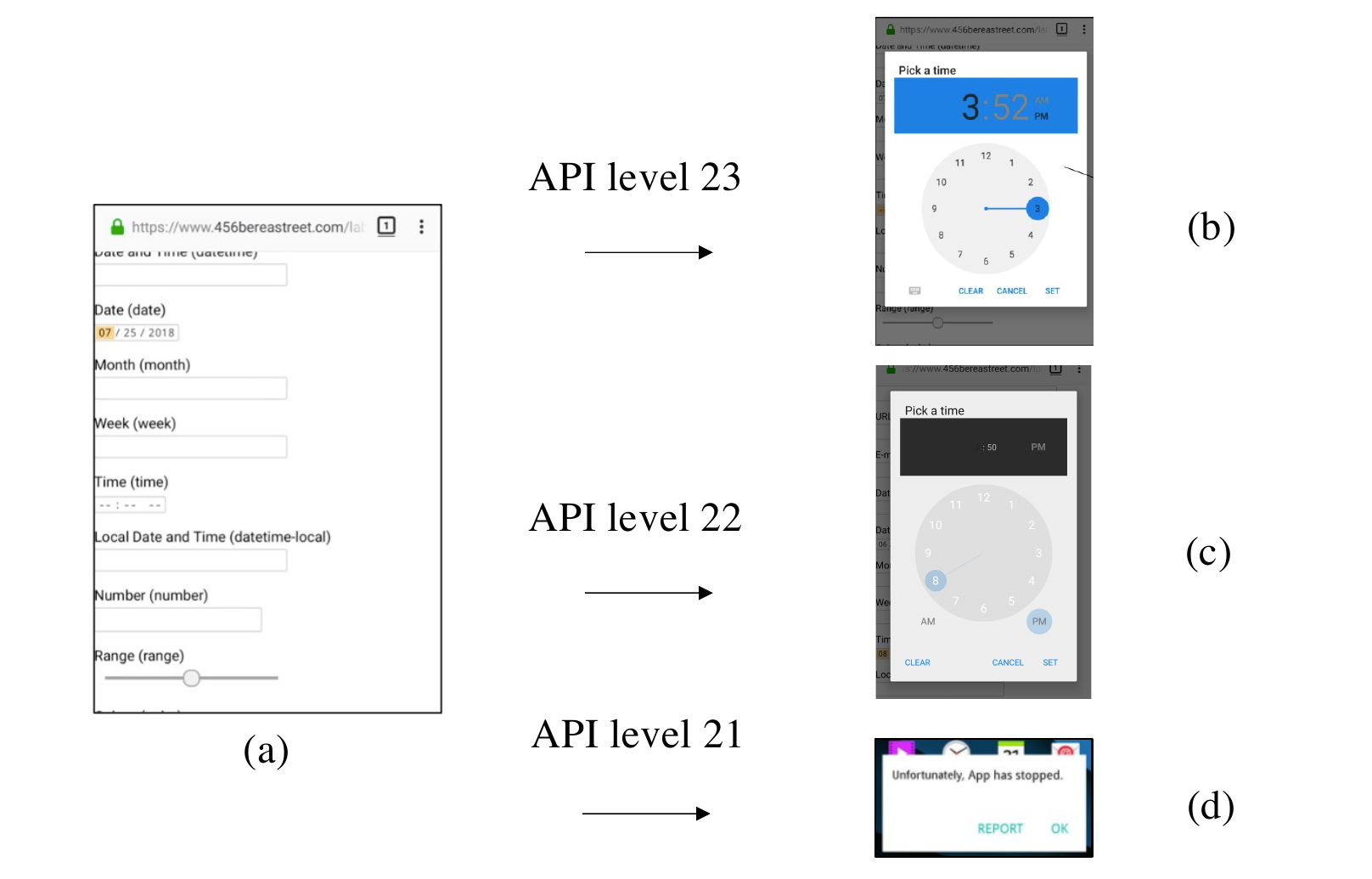}
	\caption{{Reproduction steps for the issue as shown in Figure~\ref{fig:configurationfile}.}}
	\label{fig:reproduce}
\end{figure}
It is generally difficult for developers to identify compatibility issues from many XML configuration files in their apps, especially for those issues that can only be triggered under specific conditions.
Figure~\ref{fig:reproduce} shows a test case that can manifest the issue illustrated in Figure~\ref{fig:configurationfile}.
To this end, developers should first let the app load a website containing a time picker as Figure~\ref{fig:reproduce}(a) shows, and then choose the corresponding input field to open the problematic time picker as listed in Figure~\ref{fig:configurationfile}.
It took the developers one month to detect and fix the above issue.
The issue was caused by the inconsistent processing of \texttt{android:color} in the Android framework among API levels 21, 22, and 23.
The different implementations of these API levels are shown in Figure~\ref{fig:frameworkprocess}. First, at API levels 21 and 22, the configuration API \texttt{AttributeSet\#getAttributeIntValue()} in Line 7 is invoked to load \texttt{android:color}, whose value affects the array \texttt{mSpec} in Line 8. However, there is a defect in the Android framework at API level 21. It accesses an illegal index of the array \texttt{mSpec} in Line 12, causing the app to crash. At API level 22, we did not witness the crash as the Android framework has fixed the processing of \texttt{android:color}, as shown in Line 13. 
Second, %the configuration API that loads \texttt{android:color} has been changed to \texttt{TypedArray\#getColor()} (in Line 19) at API level 23. 
the configuration API \texttt{AttributeSet\#getAttributeIntValue()} fails to load the attribute values in the style format (i.e., Line 7 of Figure~\ref{fig:frameworkprocess}). 
As a result, the style format attribute value \texttt{?android:attr/textColorSecondary} defined in \texttt{android:color} (Line 6 of Figure~\ref{fig:configurationfile}) is ignored, causing the inconsistent look-and-feel as Figures~\ref{fig:reproduce}(b) and (c) show.

The above example shows how the code changes in the Android framework induce configuration compatibility issues.
However, it is non-trivial to automatically identify such code changes.
For example, at API level 22, the statement of loading the \texttt{android:color} attribute value (Line 7 of Figure~\ref{fig:frameworkprocess}) is guarded by a condition in Line 4 as shown in Figure~\ref {fig:frameworkprocess}. An accurate analysis of the loading behavior therefore requires path-sensitivity.
However, since path-sensitive analysis is expensive, it cannot be scaled to the whole Android framework that contains a large amount of code (i.e., 4M+ LOC of API level 30) with long-term update history (i.e., 566K+ commit changes until April 2021). 
To provide insights for automatically identifying such code changes and facilitating issue detection, we conducted an empirical study (Section~\ref{sec:3}) on real-world issues to understand the common patterns of code changes that can induce configuration compatibility issues.

\section{Empirical Study}
\label{sec:3}
To facilitate automated detection of configuration compatibility issues, we conducted an empirical study on the characteristics and symptoms of such issues in real-world Android apps.
The study aims at answering the following two research questions:
\begin{itemize}
	\item \textbf{RQ1 (Issue types and root causes):} What are the common types and the corresponding root causes of configuration compatibility issues?
	\item \textbf{RQ2 (Issue symptoms):} What are the common symptoms of configuration compatibility issues?
\end{itemize}

\subsection{Dataset Collection}
We collected bug-related code revisions from well-maintained open-source
Android apps as the empirical dataset.
To this end, we searched for suitable subjects on F-Droid
~\cite{fdroid}, which is a famous repository containing high-quality open-source Android apps.
Specifically, we selected subjects that meet the following criteria: (1)
maintaining a public issue tracking system, (2) receiving more than 500 stars on
GitHub~\cite{github} (popularity), and (3) pushing the latest git commit within 
the most recent three months (well-maintenance).
We chose these three criteria because the configuration compatibility issues
located in these selected subjects are likely to affect many users due to the popularity of the apps. 
As a result, 43 apps were returned. 

In order to locate the configuration compatibility issues affecting the 43
selected apps, we used the following two types of keywords to search for
related code revisions:
\sethlcolor{orange}
\begin{itemize}
	\item Keywords related to Android framework versions. In practice, developers often indicate the specific versions of the Android framework in which compatibility issues occur in the changelog.
	Specifically, we used two keywords, \texttt{API} (API level for short), and \texttt{Android [i]} where \texttt{[i]} stands for an integer, to search for Android system versions in changelogs.
	Besides, we also looked for code revisions that contain version-specific XML files, which are stored in the path that contains a version qualifier \texttt{v[L]}, where \texttt{[L]} represents the minimum API level applicable to the files.
	\item Keywords related to XML configuration files in Android apps.
	Specifically, we chose the following two keywords: \texttt{resource}, and \texttt{AndroidManifest}, so that they can effectively cover all types of XML configuration files supported in the Android framework.
\end{itemize}
In total, 2,376 unique code revisions were identified from the 43 apps after removing duplicates from the searching results.

{Next, we conducted manual analysis on the 2,376 code revisions to refine configuration compatibility issues. Specifically, we collected the code revisions in three steps. First, we screened out the code revisions unrelated to valid configuration compatibility issues because some irrelevant code revisions (e.g., introducing new app features) can be accidentally returned by our keyword-based search. Second, we collected the incompatibility-inducing attributes and XML elements from the revision-related commit logs, bug reports, or code diffs. 
To answer RQ1, the code changes related to the incompatibility-inducing attributes and XML elements should also be identified in the update history of the Android framework to investigate how these changes can cause issues. Third, to answer RQ2, we referred to the information of code revisions and online discussions of similar issues for the consequences when developers did not handle problematic XML elements or attributes well. Eventually, we collected 196 configuration compatibility issues from code revisions as the empirical dataset.}

\subsection{RQ1: Issue Root Causes}
\label{sec:RQ1}
\begin{table}[t]
	\caption{Common root causes of configuration compatibility issues}
	\begin{tabular}{lp{5cm}r}
		\toprule
		&\multicolumn{1}{c}{\textbf{Root Causes}}         & \textbf{Issue \#} \\ \hline
		Type 1&Unavailable configuration APIs & 116 (59.2\%)       \\
		Type 2&Inconsistent configuration APIs & 42 (21.4\%)       \\
		Type 3&Inconsistent Android internal XML configuration files & 19 (9.7\%)\\
		Type 4&Inconsistent attribute dependencies    & 9 (4.6\%)        \\
		Type 5&Inconsistent attribute usages             & 9 (4.6\%)    \\
		Type 6&Inconsistent attribute default values         & 1 (0.5\%)    \\
		\bottomrule
		\label{tab:issuecategorization}
	\end{tabular}
\end{table}

We elaborated on the six common types (or causes) identified from the 196 configuration compatibility issues as shown in Table~\ref{tab:issuecategorization}.

\begin{figure}[t]
	\centering
	\includegraphics[width=0.5\textwidth]{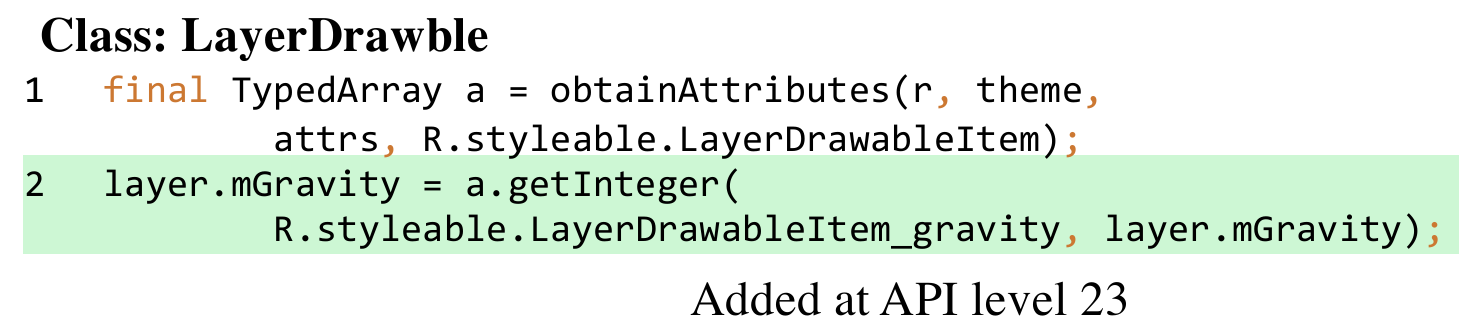}
	\caption{The Android framework code for loading the attribute value of \texttt{android:gravity} in the class \texttt{LayerDrawable}.}
	\label{fig:layerdrawable}
\end{figure}
\textbf{Unavailable configuration APIs.}
The Android framework loads attribute values by calling configuration APIs after parsing the XML tags in configuration files to \texttt{AttributeSet} or \texttt{TypedArray} objects.
Some statements invoking configuration APIs are introduced or removed as the Android framework evolves, resulting in an inability to load the associated configuration attribute values in a certain range of API levels.
In our empirical dataset, we found 116 (59.2\%) issues that were induced by such a type of code changes.
For example, the attribute value of \texttt{android:gravity} in Figure~\ref{fig:layerdrawable} is loaded by \texttt{LayerDrawable} to adjust the gravity for layer alignment starting from API level 23.
A navigation app OsmAnd~\cite{osmand} filed an issue in commit 1bbf578 that the attribute value of \texttt{android:gravity} is not loaded when running at an API level below 23, causing the incorrect display of graphic user interfaces. %There are 116 (59.2\%) issues falling into this issue type.

\textbf{Inconsistent configuration APIs.}~Configuration APIs in the Android framework are designed to load attribute values in specific data formats.
%The configuration APIs to load an attribute may vary across API levels.
Compatibility issues can happen when the configuration APIs to load an attribute vary across API levels.
The example in Figure~\ref{fig:configurationfile} between API levels 22 and 23 falls into this types.
Such an issue is caused by the style format attribute value of \texttt{android:color} not being loaded by the configuration API \texttt{getAttributeIntValue()} at API level 22.
The loading of \texttt{android:color} in an unsupported format can result in app crashes at API level 22. 
There are in total 42 (21.4\%) issues of this type.

\textbf{Inconsistent Android internal XML configuration files.}
The Android framework provides a set of internal XML configuration files that can be referenced by the developers as a part of their apps.
Compatibility issues can happen when there are changes in those internal XML configuration files as the Android framework evolves.
For example, QKSMS~\cite{qksms} commit 6b70a47 describes an issue caused by the internal XML configuration file \texttt{ic\_menu\_added.xml}, which was introduced at API level 23. There are 19 (9.7\%) issues falling into this issue type.

\begin{figure}[t]
	\centering
	\includegraphics[width=0.5\textwidth]{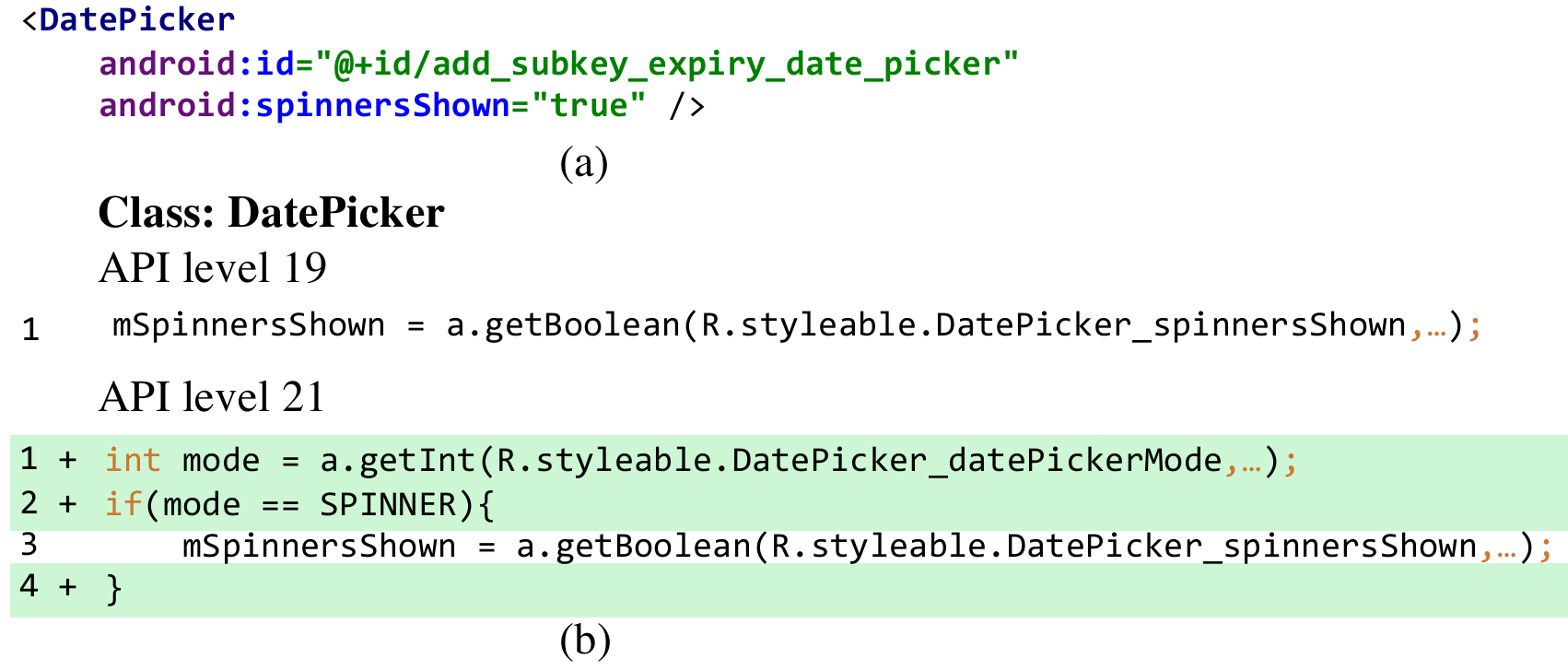}
	\caption{Code changes of loading \texttt{android:spinnersShown} in the class \texttt{DatePicker}.}
	\label{fig:datepicker}
\end{figure}
\textbf{Inconsistent attribute dependencies.}~
In the Android framework, there are dependencies across configuration attributes.
In other words, the runtime behaviors of one attribute depend on the value of other attributes.
We found nine compatibility issues that were induced by the inconsistent implementations on attribute dependencies among API levels. %of triggering conditions of invoking configuration APIs in the Android framework.
%Such changes can cause an attribute not to be loaded at specific API levels.
For example, open-keychain~\cite{openkeychain} reported an issue in commit be06c4c.
As Figure~\ref{fig:datepicker}(a) shows, developers have specified the value of \texttt{android:spinnersShown} as \texttt{true} to make the date picker widget be displayed in the spinner mode.
The attribute value cannot be loaded at API level 21 without specifying the value of \texttt{android:datePickerMode} as \texttt{spinner}, causing the date picker to be displayed in the calendar mode by default.
In this case, the app crashed at API level 21 due to a specific implementation of the date picker in the calendar mode.
The code changes in Figure~\ref{fig:datepicker}(b) show that starting from API level 21, the configuration API for \texttt{android:spinnersShown} is guarded by the conditional statement that checks whether the value of \texttt{android:datePickerMode} is \texttt{spinner}.
%Note that the existing path-insensitive analysis technique can fail to analyze the code changes falling into this type although the proportion of issues caused by such a pattern are small. 
%As the example in Figure~\ref{fig:frameworkprocess} shows, existing techniques cannot infer that \texttt{android:color} will be loaded in Line 5 without analyzing the conditional statement in Line 4. 
%Such a case can make the existing techniques derive spurious issue-inducing code changes as \texttt{android:color} is loaded in \texttt{ColorStateList} starting from API level 23.

\textbf{Inconsistent attribute usages.}
Compatibility issues can happen when there are inconsistent implementations on how the Android framework uses the attribute values after being loaded by configuration APIs.
As Figure~\ref{fig:frameworkprocess} shows, there is a change in processing the value of \texttt{android:color} (in Line 12 and 13) between API levels 21 and 22. The change avoids the \texttt{ArrayIndexOutOfBoundsException} when the Android framework parses the XML element in Figure~\ref{fig:configurationfile} at API level 22.
In total, there are nine (4.6\%) issues falling into this issue type.

\textbf{Inconsistent attribute default values.}
There is one (0.5\%) issue caused by inconsistent default values of the attribute \texttt{android:useLevel} in the XML tag \texttt{<shape>} between API levels 21 and 22, as reported in the commit a221442 of OsmAnd~\cite{osmand}.

\subsection{RQ2: Issue Symptoms}
We further analyzed the common issue symptoms as below.
Specifically, 89 (45.4\%) of the 196 issues in our empirical dataset can cause the apps to crash when triggering incompatibility-inducing XML configuration elements, as shown by the motivating example in Figure~\ref{fig:frameworkprocess}.
Another 88 issues (44.9\%) can induce an inconsistent look-and-feel across different API levels, affecting the apps' functionalities.
For example, the problem reported in the commit a221442 of OsmAnd~\cite{osmand} can force the progress bar to always show a full circle.
The remaining 19 issues (9.7\%) can cause inconsistent app behaviors beyond crashes and look-and-feel.
For example, the app Slide~\cite{slide} specified \texttt{android:requestLegacyExternalStorage} to make sure the app can still request for the external storage at an API level $\geq$ 29.
This shows that configuration compatibility issues can cause severe consequences to the app developers.

\section{ConfDroid Approach}
\label{sec:4}
In this section, we illustrate \textsc{ConfDroid}, which extracts detection rules for configuration compatibility issues. 
\textsc{ConfDroid} is built based on a novel configuration constraint that models how the attribute values specified in the XML configuration files can be processed by configuration APIs in the Android framework.
To overcome the limitations of existing static-based approaches (see Section~\ref{sec:motivating_example}), \textsc{ConfDroid} performs symbolic execution, which follows the control flows that can affect the invocations of configuration APIs, to extract configuration constraints more precisely.
Detection rules can be subsequently derived by comparing the differences of extracted configuration constraints among API levels.

\subsection{Android Configuration Constraint Model}
\label{sec:4.1}
Our empirical study found that the majority (158/196) of issues were induced by the inconsistent handling
of configuration APIs (Type 1 and Type 2) for a single attribute among API levels. The inconsistencies
can lead to \textit{failures in loading
attributes with specific data format under certain
XML tags at some API levels}.
Motivated by these findings, we formulate constraints that govern configuration
APIs, called \textbf{Android configuration constraints}.
Formally, an Android configuration constraint is defined as a tuple
$\{\mathcal{A},\mathcal{X}, \mathcal{F}\}$.
Specifically, $\mathcal{A}$ stands for the attribute to be loaded by the configuration API,
$\mathcal{X}$ is the XML tag where the attribute $\mathcal{A}$ is located; and
$\mathcal{F}$ stands for the data format that can be assigned to $\mathcal{A}$.
An Android configuration constraint is extracted when
\textsc{ConfDroid} finds a
program path in the Android framework that can invoke the configuration API to
load the attribute $\mathcal{A}$ with data format $\mathcal{F}$ under the
XML tag $\mathcal{X}$.
Rules for issue detection can then be identified by comparing the
differences in Android configuration constraints among API levels.

A constraint $\{\mathcal{A}, \mathcal{X}, \mathcal{F}\}$ for a configuration
API is extracted as follows.  $\mathcal{A}$ can be inferred from the
parameter identifying the attribute to be loaded in the configuration API call;
$\mathcal{X}$ can be inferred from the return values of the configuration API
call, such as \texttt{XmlPullParser\#getName()} in the program path (e.g., Line 4 of Figure~\ref{fig:frameworkprocess}).
$\mathcal{F}$ can be inferred from
the configuration API used to load $\mathcal{A}$ since each configuration API can only load attribute values in a specific group of data formats.
Note that one configuration API can generate multiple constraints if it can support loading attribute values in different data formats.
For example, as the code snippets in Figure~\ref{fig:frameworkprocess} show, by
analyzing the program paths to the configuration APIs
\texttt{getAttributeIntValue()} and
\texttt{getColor()}, we can obtain the following constraints
between API levels 22 and 23.
\\
\textbf{API level 22:}\\
$\{\texttt{android:color}, \texttt{item}, \texttt{int}\}\,\,\,\,\,\,\,\,\,\,\,\,\,\,\,\,\,\,\,\,\,\,\,\,\,\,\,\,\,\,\,\,\,\,\,\,\,\,\,\,\,\,\,\,\,\,\,\,\,\,\,\,\,\,\,\,\,\,\,                  (1)$\\
\textbf{API level 23:} \\
$\{\texttt{android:color}, \texttt{item}, \texttt{int}\} \,\,\,\,\,\,\,\,\,\,\,\,\,\,\,\,\,\,\,\,\,\,\,\,\,\,\,\,\,\,\,\,\,\,\,\,\,\,\,\,\,\,\,\,\,\,\,\,\,\,\,\,\,\,\,\,\,\,\,(2)$\\
$\{\texttt{android:color}, \texttt{item}, \texttt{styled\_int}\}\,\,\,\,\,\,\,\,\,\,\,\,\,\,\,\,\,\,\,\,\,\,\,\,\,\,\,\,\,\,\,\,\,\,(3)$\\

As the API descriptions in Android Developers~\cite{androiddevelopers} shows, comparing to \texttt{getAttributeIntValue()}, the configuration API \texttt{getColor()} can load integer attribute values in the style format, an additional Android configuration constraint for \texttt{android:color} (Equation 3) is extracted at API level 23.

\subsection{\textsc{ConfDroid} Overview}
We propose \textsc{ConfDroid} to identify Android configuration constraints, which can be inferred from configuration API invocations in the Android framework to model the common issue patterns (Type 1 and Type 2). 
The identified constraints can be further refined into a set of compatibility detection rules based on the differences among the identified constraints across different API levels.

To identify Android configuration constraints, we prefer using static analysis to dynamic analysis. Using dynamic analysis for the purpose is expensive due to the large search space of possible test inputs. For example, a dynamic approach needs to generate an XML file as given in Figure~\ref{fig:configurationfile} to reach the configuration API in Line 7 of Figure~\ref{fig:frameworkprocess}. The file generation involves searching for (1) the tree hierarchies of XML configuration elements (e.g., \texttt{<item>} is set as a child of \texttt{<selector>}), (2) the XML elements embedding the concerned attributes (e.g., \texttt{android:color} should only be set in \texttt{<item>}), and (3) other necessary configuration settings to avoid crash when parsing XML files (e.g., \texttt{android:state\_activated} in Line 5). In contrast, static analysis enables us to compare the approximated behaviors of these APIs without actually generating valid test inputs.
	
However, the existing techniques~\cite{rabkin2011static,behrang2015users,dong2016orplocator,chen2020understanding} analyze configuration APIs by conducting path-insensitive analysis, which can be somewhat inaccurate and miss incompatible configuration APIs without considering the if-conditions (e.g., Line 6 in Figure~\ref{fig:frameworkprocess}). Therefore, a path-sensitive analysis is necessary to obtain $\{\mathcal{A},\mathcal{X},\mathcal{F}\}$ by solving path constraints triggering configuration APIs with the constraint solver. Symbolic execution is a powerful path-sensitive analysis technique. Yet, an exhaustive symbolic execution is expensive and cannot scale to the size of the Android framework. In addition, the path constraints triggering the configuration APIs can be complex as the Android framework codebase is mixed with components compiled to native code. To tackle the above challenges, \textsc{ConfDroid} performs backward symbolic execution from the call site of configuration APIs. This is based on a trimmed version of an inter-procedural control flow graph, which is built by pruning the statements that are not needed to infer the Android configuration constraints (Section~\ref{sec:building-ICFG}). The trimmed-ICFGs are designed to not only address the scalability by reducing the number of program paths and complex conditions to be analyzed, but also preserve reasonable accuracy while conducting path-sensitive analysis. We illustrate the details of backward symbolic execution and trimmed-ICFGs in the following parts.

To summarize, \textsc{ConfDroid} works as follows.
\begin{itemize}
	\item \textsc{ConfDroid} builds \textbf{a trimmed version of interprocedural control flow graphs (trimmed-ICFGs)} for the Android framework code. We choose to build the trimmed-ICFGs for scalability (Section~\ref{sec:building-ICFG}).
	\item \textsc{ConfDroid} performs \textbf{backward symbolic execution} on the trimmed-ICFGs to extract Android configuration constraints (Section~\ref{sec:extract-configuration-constraint}).
	\item \textsc{ConfDroid} \textbf{generates detection rules} by comparing the differences of Android configuration constraints among different API levels (Section~\ref{sec:generate-rule}).
\end{itemize}

\subsection{Building Trimmed ICFG}
\label{sec:building-ICFG}
\textsc{ConfDroid} performs analysis over a trimmed version of an
inter-procedural control flow graph, which is denoted as $ICFG^T$, to ensure
the scalability.
Our idea to build such an $ICFG^T$ is inspired by the observation that the code
changes triggering compatibility issues in \textit{unavailable configuration
APIs} and \textit{inconsistent configuration APIs} primarily reside in the class that invokes the configuration APIs to load the incompatibility-inducing
attributes.
Therefore, \textsc{ConfDroid} builds an $ICFG^T$ for each class that invokes configuration APIs by combining the call graph and each method's control flow graph within the class.
The strategy limits the extraction of Android configuration constraints to
intra-class analysis. 
However, such a strategy does not pose a significant impact on the process of configuration constraint extraction and detection rule generation (Section~\ref{sec:discussion}).
%We will discuss the impact of this limitation in Section~\ref{sec:discussion}.

\subsection{Extracting Android Configuration Constraints}
\label{sec:extract-configuration-constraint}
\begin{algorithm}[t]
	\caption{Extracting Android Configuration Constraints}
	\DontPrintSemicolon
	\label{alg:1}

	\KwInput{$ICFG^T$: The trimmed-ICFG for the Android framework}
	\KwOutput{$ACC$: Android configuration constraints}
	$ACC \leftarrow []$\;
	\ForEach{$tgtStmt \in ICFG^T$ }{
		%$worklist \leftarrow \{(tgtStmt, \textsc{True})\}$\;
		add $(tgtStmt, \textsc{True})$ to $worklist$\;
		\While{\rm $worklist$ is not empty}{
			remove $(s',\phi_{post})$ from $worklist$\;
			\ForEach{\rm $s$ with an edge $(s,s')$ in $ICFG^T$}{
					$\phi_{pre} \leftarrow \phi_{pre} \vee trans(s, \phi_{post})$ \;
					\If{\rm $\phi_{pre} \neq \textsc{False}$}{
					add $(s, \phi_{pre})$ into $worklist$\;
				}
			}

	    }
		\ForEach{\rm $\pi$ of $tgtStmt$}{
			$acc \leftarrow refineACC(\pi)$\;
			add $acc$ to $ACC$\;
		}
	}
	\Return{ACC}\;
\end{algorithm}

Taking the generated trimmed-ICFGs~$ICFG^T$ as inputs,
\textsc{ConfDroid} extracts Android configuration constraints by analyzing
program paths to the target statements that contain the invocation of
configuration APIs (Algorithm~\ref{alg:1}).
Specifically, it takes the following two steps to extract Android
configuration constraints.
First, starting from each target statement $tgtStmt$, \textsc{ConfDroid} performs backward symbolic execution from the trimmed-ICFG $ICFG^T$ generated in Section~\ref{sec:building-ICFG} to extract the path constraint $\pi$ (Line 3-9).
Such a path constraint $\pi$ is a first-order logic formula that records (1) the conditions that should be satisfied to invoke the target statement, and (2) the configuration API invocation in the target statement (Line 3-9).
Unlike forward symbolic execution, backward symbolic execution saves on testing efforts by only exploring relevant program statements to reach the target statements.
Then, the path constraints will be further refined as a set of Android configuration constraints with the help of SMT solver Z3~\cite{z3} (Line 10-12).

\textbf{Extracting path constraints.}
\label{sec:extract-conf-constraint}
For each target statement $tgtStmt$ with configuration API invocations, \textsc{ConfDroid} computes the path constraints $\pi$ by performing backward symbolic execution along $ICFG^T$.
Specifically, the analysis will maintain the symbolic states $\phi_{pre}(s)$ and $\phi_{post}(s)$ representing the precondition and postcondition of $s$ in $ICFG^T$.
The symbolic states are computed iteratively along $ICFG^T$ with the symbolic state transformer $trans$, as defined in Table~\ref{tab:wp}, from $\phi_{post}(s)$~to~$\phi_{pre}(s)$.
After computing $trans(s, \phi_{post})$ for a statement $s$ and $\phi_{post}$, the algorithm merges the result with $\phi_{pre}$ that is already presented before the statement $s$ (Line 7).
In the presence of loops, we analyze them twice to ensure that the algorithm terminates and traverses the back edge of the loop at least once.
The path constraint $\pi$ is obtained from $\phi_{pre}$ of the statements $s$ that are the entry points of $ICFG^T$.

Taking the first code snippet in Figure~\ref{fig:frameworkprocess} as an
example,
\textsc{ConfDroid} will analyze the target statement in Line 7, and compute its path constraint $\pi$ along the path (7,6,5,4,3) as follows:

\begin{itemize}
	\item $\phi_{pre}$ in Line 7, $tgtStmt=attrs.getAttributeIntValue\\(i,color)$;
	\item $\phi_{pre}$ in Line 6, $tgtStmt = attrs.getAttributeIntValue\\(i,\,color) \wedge i = R.attr.color$;
	\item $\pi = \phi_{pre}$ in Line 3 (entry point), $tgtStmt = getXml().getAttributeIntValue(i,\,color) \wedge  i = R.attr.color \wedge getXml().getName().equals(``item")$;
\end{itemize}

\begin{table}[t]
	  \begin{threeparttable}[b]
		\caption{Specification of symbolic state transformer $trans$ for Java statements.}
	\label{tab:wp}
	\begin{tabular}{p{3.5cm}|p{4.5cm}}
		\toprule
		$statement$ & $trans(statement, \phi)$ \\ \midrule
		$tgtStmt$  & $\phi \wedge tgtStmt$ \\
		$x = op\, y$& $\phi[y/x]$     \\
		$x = y\, op \, z$ & $\phi[(y \, op \, z)/ x]$      \\
		$x = api(y_1, ..., y_n)$&  $\phi[api(y_1, ..., y_n)/ x]$    \\
		$x.y = z$&   $\phi[z/x.y]$   \\
		$arr[i] = x$&$\phi[x/arr[i]]$ \\
		\textbf{if} $c$ (branch condition) & $\phi \wedge c$ \\
		\bottomrule
	\end{tabular}
   \begin{tablenotes}
	\item $*$ The string APIs modelled by Z3 (e.g., \texttt{String\#equals()}) are considered as $op$ instead of $api$.
	\item $*$ $\phi[x/y]$ means replacing the symbolic variable $x$ in the symbolic state $\phi$ by $y$.
	\end{tablenotes}
	\end{threeparttable}
\end{table}

\textbf{Refining Android configuration constraints.}
\textsc{ConfDroid} further refines the path constraints $\pi$ as Android configuration constraints $acc$ with the help of SMT solver Z3~\cite{z3} (Line 10-12).
First, \textsc{ConfDroid} checks whether the target statements are reachable by the path constraints.
To achieve this, \textsc{ConfDroid} substitutes API calls in the path constraints to symbolic variables according to the return types.
\textsc{ConfDroid} will declare any variables whose data formats are not modeled in Z3 as integer variables and replace constants (e.g., \texttt{null}) in the path constraints as integers in queries of Z3.
The target statement is reachable if its associated path constraint is decided to be satisfiable by Z3.

\textsc{ConfDroid} then calculates $\mathcal{A}$ and $\mathcal{X}$ from the path constraints $\pi$.
$\mathcal{A}$ is calculated from the parameter identifying the attribute to be loaded in the API call (i.e., $i$ in Line 7 of Figure~\ref{fig:frameworkprocess}). 
$\mathcal{X}$ is calculated as the return value of the API call \texttt{XmlPullParser\#getName()}. 
Note that $\mathcal{X}$ will be assigned as the class name where $\pi$ locates when the concrete value of $\mathcal{X}$ cannot be identified, because the Android framework allows app developers using the class names as XML tags.
\textsc{ConfDroid} then leverages Z3 to infer all the possible values of the above symbolic variables for $\mathcal{A}$ and $\mathcal{X}$.
We set the time budget as one minute to Z3 for each symbolic variable.
\textsc{ConfDroid} discards the cases when (1) the value of symbolic variables for $\mathcal{A}$ or $\mathcal{X}$ are undecidable, or (2) \textsc{ConfDroid} cannot obtain all possible values of symbolic variables for $\mathcal{A}$ or $\mathcal{X}$ within the time budget.
Such a strategy does not pose significant problems to the accuracy of Android configuration constraint extraction. Since those discarded cases only account for 1.8\% (29/1534) when applied to API level 30 (Section~\ref{sec:discussion}).

Next, to infer the value of $\mathcal{F}$, we manually built a map between each configuration API and its supported data format based on its API descriptions. \textsc{ConfDroid} further uses the map of configuration APIs with their supported data formats to analyze $\mathcal{F}$ for $tgtStmt$.
Finally, \textsc{ConfDroid} extracts Android configuration constraints $acc$ with all the possible combinations of $\mathcal{A}$, $\mathcal{X}$ and $\mathcal{F}$.

\subsection{Generating Detection Rules}
\label{sec:generate-rule}
Detection rules are further inferred by comparing the differences in Android configuration constraints between the two adjacent API levels $l_1$ and $l_2$ ($l_1<l_2$) as follows.
\begin{itemize}
	\item \textsc{ConfDroid} reports a rule of \textit{attribute loading change} if an attribute $\mathcal{A}$ of the XML tag $\mathcal{X}$ can be loaded at the API level $l_1$ only but not $l_2$ (or at $l_2$ only but not $l_1$).
	The rules falling into this type are induced by unavailable configuration APIs (Type 1 in Section \ref{sec:RQ1}).
	\item \textsc{ConfDroid} reports a rule of \textit{data format change} if there are inconsistencies in the supported data formats $\mathcal{F}$ for an attribute $\mathcal{A}$ of the XML tag $\mathcal{X}$ between $l_1$ and $l_2$.
	The rules falling into this type are induced by inconsistent configuration APIs (Type 2 in Section \ref{sec:RQ1}).
\end{itemize}

For example, by comparing the differences of Android configuration constraints as shown in Section \ref{sec:4.1}, 
\textsc{ConfDroid} generates a rule of data format change as follows.
$$ \left\{
\begin{aligned}
	\textbf{Attribute:\,\,} & \texttt{android:color}&
	\textbf{XML tag:\,\,} & \texttt{item} \\
	\textbf{Data format:\,\,} & \texttt{styled\_int}&
	\textbf{API level:\,\,} & [22,23] \\
\end{aligned}
\right\}.
$$

The above rule indicates that assigning the attribute value in the styled integer format to
\texttt{android:color} under the XML tag \texttt{<item>} can trigger
compatibility issues between API levels 22 and 23.

\subsection{Discussion}
\label{sec:discussion}
The accuracy of detection rules extracted by \textsc{ConfDroid} can be affected by the limitations of conducting backward symbolic execution to infer Android configuration constraints.
First, \textsc{ConfDroid} performs an intra-class level backward symbolic execution, which can cause $\mathcal{A}$ and $\mathcal{X}$ in Android configuration constraints identified inaccurately.
Second, \textsc{ConfDroid} discards cases where  values of $\mathcal{A}$ and $\mathcal{X}$ cannot be inferred due to the inabilities of Z3.
Therefore, false positive rules occur when an Android configuration constraint cannot find the equivalent one at other API levels due to the limitations of our analysis.
Plus, false negative rules are incurred when \textsc{ConfDroid} fails to generate Android configuration constraints for the evolved attributes.
However, \textsc{ConfDroid} only discards 1.9\% (29/1534) of the configuration API
call sites at API level 30, indicating the
insignificant impact on the inference of Android configuration constraints due
to the intra-class level symbolic execution and the inabilities of Z3.
Although we cannot draw a whole picture of all the cases of Android
configuration evolutions, we find that \textsc{ConfDroid} is only unable to
generate Android configuration constraints for 21.0\% (301/1435) attributes at
API level 30. These constraints are accountable for the false negatives in the
warnings reported by \textsc{ConfDroid} (Section~\ref{sec:rq3}).

%We measured the impact of the above limitations as follows.
%First, from the 1,505 configuration APIs that were successfully handled by \textsc{ConfDroid} at API level 30, we performed manual validation on 200 randomly sampled configuration APIs regarding 488 Android configuration constraints.
%An Android configuration constraint is considered valid if we can find online
%references to show the attribute $\mathcal{A}$ with data
%format $\mathcal{F}$ can be specified in the XML tag $\mathcal{X}$.
%We only failed to validate four Android configuration constraints regarding two configuration APIs.
\section{Evaluation}
\label{sec:5}
We implemented \textsc{ConfDroid} based on Soot~\cite{lam2011soot}.
In our evaluation, we answer the following research questions:
\begin{itemize}
	\item \textbf{RQ3 (Effectiveness of detection rule extraction):} What is the effectiveness of \textsc{ConfDroid} on extracting detection rules compared with baseline methods?
	\item \textbf{RQ4 (Usefulness):} Can rules that are uniquely extracted by \textsc{ConfDroid} be useful for detecting issues in real-world Android apps?
\end{itemize}

\subsection{RQ3: Effectiveness of Detection Rule Extraction}
\label{sec:rq3}

To answer RQ3, we ran \textsc{ConfDroid} on the Android framework code among
API levels between 21 and 30 (inclusive). We evaluated its accuracy in
extracting the detection rules of configuration compatibility issues.
We compared the results with the following baseline methods.
\begin{itemize}
	\item \textbf{Baseline I:} \textsc{Lint}~\cite{lint}, the popular static analyzer officially released by Google.
	By modeling the API levels when attributes were introduced, \textsc{Lint} can detect configuration compatibility issues caused by an unavailable attribute $\mathcal{A}$ at API level $\mathcal{L}$.
	\item \textbf{Baseline II:}
	\textsc{ORPLocator}~\cite{dong2016orplocator}, the state-of-the-art
	technique for extracting attributes in a software system by conducting path-insensitive analysis on its configuration APIs.
	We ran \textsc{ORPLocator} at different API levels to extract detection
	rules for unavailable attribute $\mathcal{A}$ at API level $\mathcal{L}$.
	\item \textbf{Baseline III:}
	\textsc{cDep}~\cite{chen2020understanding}, the state-of-the-art software misconfiguration tool which focuses on analyzing attribute dependencies in the software system.
	We ran \textsc{cDep} among API levels to generate detection rules for the unavailable dependencies between two attributes $\mathcal{A}_1$ and $\mathcal{A}_2$ at API level $\mathcal{L}$.
\end{itemize}

Note that we did not choose \textsc{SCIC}~\cite{behrang2015users}, the state-of-the-art tool for detecting misconfigurations caused by software evolution, as the baseline since its artifact is not available. Besides,
\textsc{ORPLocator} (Baseline II) and \textsc{cDep} (Baseline III) are more recently published techniques.
The original algorithms of \textsc{ORPLocator} and \textsc{cDep} comprise
inter-class analyses. However, they cannot scale up to the Android framework
and fail to extract any rules. 
As such, we slightly adapted \textsc{ORPLocator} and \textsc{cDep} to
performing intra-class analyses in the construction of the two baselines.
%\scc{Please check if my rewriting is correct.}
%\huaxun{Yes.}

\textbf{Validating detection rules.}
We considered an issue-detection rule $r$ as valid if the XML configuration
files that satisfy the triggering conditions specified by $r$ can induce
configuration compatibility issues between two API levels.
%We followed the strategies conducted by an existing study\cite{wei2019pivot} to validate rules.
%Note that the issue-detection rules produced by different methods take different forms.
Specifically, for each issue-detection rule $r$ extracted by \textsc{ConfDroid} and other baselines,
we used the attribute name, the XML tag, and the incompatibility-inducing API levels
specified in $r$ as keywords to search for relevant discussions in Stack
Overflow~\cite{stackoverflow} and GitHub~\cite{github}.
For those rules $r$ with no relevant discussions found, we manually crafted an
Android app that includes XML configuration files satisfying the
issue-triggering condition in $r$ to manifest inconsistent behaviors
across API levels.
Furthermore, the manifested inconsistent behaviors should disappear after
removing the incompatibility-inducing attribute from the concerned XML configuration file.
A rule $r$ is considered valid if we can find a relevant online discussion or
successfully craft an app confirming the existence of configuration
compatibility issues.

\begin{table}[t]
	\caption{Execution Time for Rule Extraction and Validation Rate of
	Extracted Rules. Rules in Lint are hardcoded.}
	\centering
	\begin{center}
	\begin{tabular}{p{40pt}p{40pt}p{30pt}p{40pt}p{35pt}l}
		\toprule
		\textbf{Method} {\centering}   &  {\centering \textsc{ConfDroid}} &  {\centering \textsc{Lint}}  & \textsc{ORPLocator} {\centering} &{ \centering \textsc{cDep}} \\ \hline
				 		\textbf{Time}  &  2h58m36s  & - & 2m26s & 3m25s  \\\hline
		 \textbf{Validation Rate}  &   91.9\% (282/307)  & 100.0\% (218/218) &
		 85.6\% (131/153) & 12.7\% (28/220) \\

\bottomrule& & &
\label{tab:ruleextraction}	\end{tabular}
\end{center}
\end{table}

\begin{comment}
	\begin{figure}[t]
	\centering
	\includegraphics[width=0.28\textwidth]{./img/issue-distribution.pdf}
	\caption{Valid issue-detection rules generated by \textsc{ConfDroid} and the other two baselines. \civi{Make the font in the Figure larger}}
	\label{fig:issuedistribution}
	\end{figure}
\end{comment}

Table~\ref{tab:ruleextraction} shows the execution time as well as the number
of rules
extracted by \textsc{ConfDroid} and baseline methods.
The execution time of \textsc{Lint} is inapplicable as the detection rule is
hardcoded.
Although the execution time of \textsc{ConfDroid} is larger than
path-insensitive approaches \textsc{ORPLocator} and \textsc{cDep},
\textsc{ConfDroid} can extract more valid detection rules and complete
analyzing the Android frameworks within three hours.
In total, \textsc{ConfDroid} successfully extracts 282 valid detection rules,
achieving a precision of 91.9\%.
By performing backward symbolic execution on the configuration APIs in the
Android framework, 
the 65 unique valid detection rules (47 unavailable configuration
APIs and 18 inconsistent configuration APIs) extracted by \textsc{ConfDroid}
are not found in any of the three baselines.

On the other hand, the 25 detection rules extracted by \textsc{ConfDroid} cannot be validated due to the following two reasons.
First, the intra-class level symbolic execution and the inabilities of using Z3 to process configuration APIs can cause incomplete configuration constraint extraction.
Second, the Android framework has proposed an internal workaround at the stage of the Android app compilations,
which is not analyzed by \textsc{ConfDroid}.
Besides, 14.4\% of the rules extracted by \textsc{ORPLocator} are invalid due to
(1) the adaptation of intra-class level analysis to ensure scalability, and (2) the inaccuracy of its path-insensitive analysis on the Android framework.
While we have resolved the scalability issues of the original version of
\textsc{cDep}, it achieves the lowest rule validation rate because of
(1) extracting incomplete attribute dependencies by only performing intra-class
level data flow analysis, and (2) identifying minor code changes that will not
induce compatibility issues.

Still, we found 28 rules (28/218=12.8\%) that are uniquely included in \textsc{Lint} but not \textsc{ConfDroid}.
The false negative rules of \textsc{ConfDroid} concern the theme attributes (e.g., \texttt{android:attr/textColorSecondary} in Figure~\ref{fig:configurationfile}) whose value can only be referenced by another attributes in XML.
Such theme attributes are not loaded using the Android configuration APIs and
hence are missed by \textsc{ConfDroid}.
We have not witnessed any false negative rules in \textsc{ConfDroid} that were
extracted by \textsc{ORPLocator} because the valid rules of 
\textsc{ConfDroid} and \textsc{ORPLocator} fall into the pattern of \textit{unavailable
configuration APIs}.
We also did not witnessed any false negative rules that are extracted by
\textsc{cDep}. 
The \textsc{cDep} tool is designed based on the change pattern in attribute dependencies (Type 4 in Section~\ref{sec:RQ1}).
Such a pattern is not modelled by \textsc{ConfDroid} since it only accounts for 4.6\% of the issues in the empirical study dataset.
%Besides, more than 80\% of the compatibility issues were induced by a single attribute in our empirical study.
In fact, the validation rules generated by \textsc{cDep} are all caused by the introduction and removal of attributes, which are also dependent on other attributes in the Android framework.
Moreover, false negatives can also be induced by the inabilities of Z3 in
processing configuration APIs.
Although it is challenging to measure false negatives as we lack an authoritative
ground truth, we argue that the number of false negatives can be small since \textsc{ConfDroid} can
successfully process 98.1\% (1504/1533) of configuration APIs, which cover
79.0\% (1134/1435) of public attributes at the API level 30.

\subsection{RQ4: Usefulness}
\sethlcolor{green}
\begin{table}[t]
	\caption{{Number of warnings detected using the rules extracted by
	\textsc{ConfDroid} and baselines}}
	\centering
\begin{center}
		\begin{tabular}{lcccc}
\toprule
 &  {\centering \textsc{ConfDroid}} &  {\centering \textsc{Lint}}  & \textsc{ORPLocator} {\centering} & \textsc{cDep} \\ \hline
$D$&50,157&9,998&8,424&4,304\\
$F_{v}$ & 46,580  & 7,316 & 5,433 & 3,595 \\
$F_{lib}$ & 41,790 & 5,007 & 4,134 & 4,290 \\
$D-F_{v}-F_{lib}$ & 688 & 435 &  54& 140 \\
 \bottomrule
\label{tab:issuedetection}	\end{tabular}

\thefootnote{\leftline{$D$: Warnings generated by rules only.}}
\thefootnote{\leftline{$F_{v}$: False warnings identified by API level filter.}}
\thefootnote{\leftline{$F_{lib}$: False warnings identified by support library filter.}}
\thefootnote{\leftline{$D-F_{v}-F_{lib}$: The final output after filtering false warnings by $F_{v}$ and $F_{lib}$}}
\end{center}

\end{table}

To answer RQ4, we checked whether the rules of \textsc{ConfDroid} could be applied to detect real issues with the comparison of three baselines. Note that Android has introduced various protection mechanisms documented in Android Developers~\cite{androiddevelopers} to prevent configuration compatibility issues. The protection mechanisms include (1) checking the API level ranges in which the XML configuration file can be used and (2) using the Android support library (AndroidX) to parse XML configuration files. Leveraging detection rules alone without considering any protection mechanisms can induce a large number of false warnings. Therefore, we built two false warning filters $F_{v}$ and $F_{lib}$ to investigate the impact of protection mechanisms on issue detection. Specifically, $F_{v}$ checks whether the XML configuration file can be applied to incompatibility-inducing API levels. App developers can provide different copies of the XML configuration files for different API levels with the identifier \texttt{-v} (e.g., \texttt{-v24} in \texttt{layout-v24/a.xml}). For an issue reported by a rule $r$ with problematic API levels $[l_1, l_2]$, $F_{v}$ considers the API level identifier with the parameter value of \texttt{minSdkVersion} to check whether the XML file can be used at $l_1$. $F_{v}$ also checks whether there is another copy of the file for API level $l_2$. The reported issue is considered as invalid if the XML file cannot be used in both $l_1$ and $l_2$. $F_{lib}$ checks whether an XML configuration file can be parsed by the Android support library (AndroidX) instead. Here, we consider an XML configuration file to be free of compatibility issues if it is only referenced by the APIs or attributes defined in the library. We then built four different variants of issue detectors with two filters $F_{v}$ and $F_{lib}$ as Table~\ref{tab:issuedetection} shows.
	
We applied the above variants to detect issues in real-world Android apps.
We crawled 116 open-source Android apps in F-Droid~\cite{fdroid} with at least 50 stars in GitHub~\cite{github} and the last git commit made within the past six
months.
We further collected 200 closed-source apps marked as top-ranked in AppBrain~\cite{appbrain}.
None of these 316 apps overlaps with those used in our
empirical study (Section~\ref{sec:3}) to avoid the risk of overfitting.

\sethlcolor{green}{Table~\ref{tab:issuedetection} shows the number of warnings produced by each variant of issue detectors. We can see that the number of false warnings is significantly less when combining two false warning filters $F_{v}$ and $F_{lib}$, showing that the above two mechanisms are often adopted to handle compatibility issues. Several frequently-used attributes contribute to the number of warnings reported in Table~\ref{tab:issuedetection}. For example, developers intensively use an incompatibility-inducing attribute \texttt{android:strokeColor} in their apps. $F_{v}$ helps reduce 12,445 false warnings that developers have properly used such an attribute with API level identifiers. }

Considering the output of $D-F_{v}-F_{lib}$, the rules of \textsc{ConfDroid} generated 688 issue warnings, outperforming the other baseline methods.
Note that the number of warnings can be varied depending on the frequency of which incompatibility-inducing attributes are used in apps.
%Among them, 218 warnings were generated on 33 open-source apps in F-Droid, and 517 warnings were generated on 98 closed-source apps.
%Although 31 valid rules can be extracted by
%\textsc{Lint} and \textsc{cDep} but not \textsc{ConfDroid}, we only witnessed two warnings generated by them as the incompatibility-inducing attributes concerning those rules were not widely used by app developers.
As for the 65 rules that were uniquely identified by \textsc{ConfDroid}, 11 of them were activated to output 253 warnings from 74 apps, among which 90 warnings were from 20 open-source apps, while the remaining 163 warnings were from 56 closed-source apps.
The above results show that the rules uniquely generated by \textsc{ConfDroid} can be leveraged to generate issue reports that are unknown to other baselines.

We then conducted manual reproduction trying to manifest inconsistent runtime behaviors on the warnings that are uniquely reported by rules in \textsc{ConfDroid}.
Specifically, we first tried to build test cases to reach the \texttt{Activity} using the XML element in the issue reports.
Then, we tried to manifest the runtime behavior controlled by the incompatibility-inducing attributes in the XML element by reading the code logic of the \texttt{Activity}.
Since we have no access to the source code of the closed-source apps, we
analyzed the apps' code logic by reading the smali code decompressed from their
apk files.
An issue is considered reproducible if we can observe inconsistent runtime
behavior by building a test case based on the above two steps.
We successfully reproduced 107 warnings in 30 apps (67 warnings in 12 open-source apps and 40 warnings in 18 closed-source apps).
We failed to reproduce the remaining warnings for the following three reasons.
First, the XML elements where the incompatibility-inducing attributes are located were not reachable because of special triggering conditions (e.g., paying for unlocking hidden functionalities).
Second, false warnings were generated because it cannot recover the
workaround of configuration compatibility issues made by app developers.
Third, the app was obfuscated so that it is difficult for us to understand the app's runtime behavior through analyzing the smali code.
\sethlcolor{lightgray}
\begin{table}[t]
	\caption{{Issue Reports}}
	\centering
	\begin{tabular}{cccccl}
		\toprule
		 {\textbf{App}}   & \textbf{ID} {\centering} & \textbf{Warnings} & \textbf{Confirmed} & \textbf{Fixed} \\ \hline
		AndOTP~\cite{andotp}  & 539$^1$ & 42&42 & 42 \\
		Aria2App~\cite{arial2app} &42& 2& 0& 0\\
		Dash Wallet~\cite{dashwallet} &648& 3&0 &0 \\
		document-viewer~\cite{documentviewer} &328&1 &0 &0 \\
		FastScroll~\cite{androidfastscroll} &36$^1$& 1&1 & 0\\
		GoodTime~\cite{goodtime}  & 197$^1$ &1 & 1 & 1 \\
		Metro~\cite{metro} &55& 6& 0& 0\\
		Kontalk~\cite{kontalk} &1315& 1&0 &0 \\
		openWorkout~\cite{openworkout} &45&1 &0 &0 \\ 
		PersianCalendar~\cite{droidpersiancalendar}  & 621$^1$ &5 & 5 & 5 \\
		Twire~\cite{twire}  & 119$^1$ & 3& 3&3 \\
		UniPatcher~\cite{unipatcher} &52&1 &0 &0 \\ \hline
		\textbf{Total} &-&67 &52 &51 \\
		\bottomrule& 
		\label{tab:issuereport}& & 	\end{tabular}
	\footnote{*}{$^1$Issues that have received replies from app developers.}

\end{table}

We further reported the 67 reproducible warnings concerning 12 open-source apps
to the original app developers.
For each app, we only sent one issue report containing all warnings found in it.
Table~\ref{tab:issuereport} shows the issue reports that have been sent to the developers.
The distribution of reported issue warnings depends on the attribute usages in different apps. For example, our issue detector reports 42 warnings in the evaluation subject AndOTP since it frequently uses the attribute \texttt{android:fillColor} to define the colors in the app icons. The processing of \texttt{android:fillColor} in the Android framework has been changed between API levels 23 and 24, causing an inconsistent look-and-feel on the app's icons.
So far, 52 warnings concerning five apps have been confirmed, and 51 warnings in four apps have been fixed by developers.
We did not report issues for closed-source apps as they do not have a public issue tracker.
Instead, we released the reproduction results of both open-source and closed-source apps on our project website~\cite{confdroid}.
The above confirmed or fixed warnings demonstrate the usefulness of rules extracted by \textsc{ConfDroid} to facilitate configuration compatibility issue detection.

\section{Threats to Validity}

{\textbf{Keywords for dataset collection.} When collecting configuration compatibility issues, we used two sets of keywords related to API levels and XML files. Although different apps may use different app-specific terms to refer to the same concepts,   such app-specific terms are hard to collect. Besides, using app-specific terms as keywords can return many irrelevant results. Therefore, our dataset was collected primarily using the above two groups of general keywords, which retrieves 196 configuration compatibility issues.
}

\textbf{Generality of our empirical findings.}
In this paper, we studied the configuration compatibility issues in the Android apps and further proposed \textsc{ConfDroid} to facilitate automatic issue detection.
Our empirical findings may not be generalized to other types of software systems.
We chose Android as our study platform because it is representative of a popular system supporting a high degree of configurability in thousands of attributes.
The open-source Android framework code provides a nice foundation for us to
study the root causes of configuration compatibility issues.

\textbf{Empirical subject selection.}
%We selected  as our empirical dataset to study
%configuration compatibility issues.
The findings from the issues in the open-source Android app can be biased to the subjects selected in our empirical study.
We mitigated the threat by selecting a significant number of issues and apps.
As a result, the study was based on 196 issues from 43 open source apps that are popular and well-maintained.

\textbf{Evolving Android framework versions.} The Android API
levels that we studied may become obsolete over time. It is a threat
common to most studies based on Android. To mitigate the
threat, we base our study on the latest API levels.

{\textbf{Manual inspections and developers' feedbacks.} We manually inspected our evaluation to validate issues detected by \textsc{ConfDroid}. The manual process can be subject to errors. We also submitted the validated ones to the original app developers for their feedback. 
However, several issue reports are still pending. This is most likely because some developers tend to be less responsive to the issue reports in the issue tracker. To address the threat that our reported issues may not be real ones, our manual process involves validating detection rules and issuing warnings through actual executions. We also make our validation steps and artifacts publicly available~\cite{confdroid}.}

\section{Related Work}

\subsection{Software Misconfigurations}
There are plenty of works~\cite{rabkin2011static, xu2013not
,behrang2015users,xu2016early, dong2016orplocator, chen2020understanding,
toman2016staccato} on detecting software misconfigurations.
For example,
Behrang et al.~\cite{behrang2015users} proposed  \textsc{SCIC} to study the attributes that cannot take effects because of software evolution.
%Xu et al.\cite{xu2016early} proposed \textsc{PCheck} that analyzes source code for early detection of latent configuration errors.
Dong et al.~\cite{dong2016orplocator} proposed \textsc{ORPLocator} to detect
inconsistencies between documentations and configuration attributes in system
code.
Recently, Chen et al.~\cite{chen2020understanding} proposed
\textsc{cDep} to extract attribute dependencies from source code.
Although these approaches can be adapted to detect compatibility issues
across multiple system versions, the proposed path-insensitive analyses are ineffective for the Android framework. It is because the loading of configuration attributes in the Android framework is commonly guarded
by variables defined in the same configuration class. In contrast,
\textsc{ConfDroid} conducts intra-class level path-sensitive analysis for
more precise detection of compatibility issues.

There are works studying misconfigurations caused by system evolution~\cite{zhang2014configuration, zhang2021evolutionary}.
For example, Zhang et al.~\cite{zhang2014configuration} proposed a technique for debugging software misconfigurations induced by software evolution.
However, the technique requires developers to provide test cases to manifest
such issues, limiting its applicability.
Recently, Zhang et al.~\cite{zhang2021evolutionary} conducted an empirical study to understand how the configuration design and implementation evolve in cloud systems.
The study does not provide a technique for issue detection.
\textsc{ConfDroid} fills the gap by incorporating automated techniques to
extract issue-detection rules, which can be readily deployed at \textsc{Lint}
for issue detection.

There are also a few related works~\cite{attariyan2012x,
rabkin2011precomputing, zhang2013automated, zhang2015proactive } on automated
software misconfiguration diagnosis.
However, their research goal is to help understand the root causes of known
software misconfigurations. Their application scenario differs from
that of \textsc{ConfDroid} where the misconfigurations are unknown.

\subsection{Android Compatibility Issues}
A few studies have been conducted on Android compatibility issues.
For example, Wei et al.~\cite{wei2016taming,wei2018understanding} studied the root causes, symptoms, and fixing
practices of Android compatibility issues.
Huang et al.~\cite{huang2018understanding} studied compatibility issues in Android callback APIs.
Hu et al.~\cite{hu2018tale} found Android compatibility issues in the WebView component.
Cai et al.~\cite{cai2019large} studied the compatibility issues occurring at
installation time and runtime.
Xia et al.~\cite{xiaandroid} studied the practice of app developers to
handle Android compatibility issues.
These studies do not cover the common root causes and patterns of Android
configuration compatibility issues, which are prevalent and can cause severe
consequences in Android apps.

Researchers have proposed a set of tools to detect Android compatibility issues.
First, dynamic-based approaches~\cite{fazzini2017automated,ki2019mimic} can
generate tests to identify app GUI inconsistencies across Android devices.
These approaches adopt a random test generation strategy, which is ineffective in triggering the inconsistencies within the huge search space of an
app's configuration attributes.
Second, existing static-based approaches~\cite{wei2016taming,wei2019pivot,huang2018understanding,li2018cid,li2018elegant,he2018understanding,wei2018understanding} take a set of predefined patterns to facilitate compatibility issue detection in Android apps.
However, their approaches can only be leveraged to detect issues caused by
problematic API invocations. These approaches are inapplicable for the detection
of configuration compatibility issues in Android apps.
Besides, \textsc{Lint}~\cite{lint}, a popuplar static-based bug detection tool
in the industry~\cite{wei2017oasis}, can only detect configuration compatibility
issues due to the introduction of new attributes.
In comparison, \textsc{ConfDroid} encodes common issue patterns to extract detection rules from the Android framework.
The evaluation results show that such rules can detect issues that were
previously unknown to app developers.

\section{Conclusion \& Future Work}
%Configuration compatibility issues in Android apps are prevalent and can cause severe consequences.
To help app developers tackle configuration compatibility issues, in this study, we first collected 196 configuration compatibility issues to empirically understand the common root causes and patterns of such issues.
Based on the findings, we further proposed \textsc{ConfDroid}, which encodes common issue patterns to automatically extract detection rules for configuration compatibility issues.
The results show that the rules extracted by \textsc{ConfDroid} can facilitate detecting issues that were previously unknown to app developers.
In the future, we plan to investigate automated repair techniques for configuration compatibility issues in Android apps.
We also plan to design a technique that can facilitate automated validation of configuration compatibility issue to minimise manual efforts.

\section*{Acknowledgment}
We sincerely thank anonymous reviewers for their valuable comments.
This work was supported by the National Natural Science Foundation of China (Grant No. 61932021, No. 61802164 and No. 62002125), Hong Kong RGC/GRF (Grant No. 16211919), Hong Kong RGC/RIF (Grant No. R5034-18), and Guangdong Basic and Applied Basic Research Foundation (Grant No. 2021A1515011562). Huaxun Huang was supported by the Hong Kong PhD Fellowship Scheme. Lili Wei was supported by the Hong Kong RGC Postdoctoral Fellowship Scheme.

%\balance
\bibliographystyle{IEEEtran}
\bibliography{sample-base}

\end{document}